\documentclass[preprint,review,12pt]{elsarticle}
\usepackage{times}
\usepackage{amsmath,amssymb}
\usepackage[usenames,dvipsnames]{xcolor}
\usepackage[colorlinks=true,linkcolor=Red,citecolor=Green,linktoc=page]{hyperref}
\usepackage{color}

\newcommand{\removed}[1]{}
\newcommand{\bE}{\mathbf{E}}
\newcommand{\be}{\mathbf{e}}
\newcommand{\bk}{\mathbf{k}}
\newcommand{\bq}{\mathbf{q}}

\newcommand{\bp}{\mathbf{p}}
\newcommand{\bv}{\mathbf{v}}

\journal{Journal of Magnetism and Magnetic Materials}
\begin{document}
\begin{frontmatter}
\title{Electron drag in ferromagnetic structures separated by an insulating interface}

\author{V. I. Kozub\footnote{Corresponding author, Email: \texttt{ven.kozub@mail.ioffe.ru}}}
\author {\fbox{M. I. Muradov}}
\address{A.~F.~Ioffe Physico-Technical Institute of Russian
Academy of Sciences, 194021 St. Petersburg, Russia}

\author{Y. M. Galperin}
\address{Department of Physics, University of Oslo, 0316 Oslo, Norway and
A.~F.~Ioffe Physico-Technical Institute of Russian
Academy of Sciences, 194021 St. Petersburg, Russia}

\date{Revisited November 1, 2017}

\begin{abstract}

We consider electron drag in a system  of two ferromagnetic layers separated by an insulating interface. The source of it is expected to be  magnon-electron interactions. Namely, we assume that the external voltage is applied to the ``active" layer stimulating electric current through this layer. In its turn, the scattering of the current-carrying electrons by magnons leads to a magnon drag current within this layer. The  3-magnons interactions between magnons in the two layers (being of non-local nature) lead to magnon drag within the ``passive" layer which, correspondingly, produce electron drag current via processes of magnon-electron scattering. We estimate the drag current and compare it to the phonon-induced one.
\end{abstract}


\end{frontmatter}

\section{Introduction}

It is of no doubt that the ferromagnetic elements play important role in modern electronics, in particular, in memory devices. The knowledge of details related to processes of switching in such devices is, naturally, of great importance. Such processes inevitably involve dynamical properties of ferromagnets. In its turn, one expects that the elementary excitations in ferromagnets affect these properties. The most known elementary excitations of magnetic nature in ferromagnets are, naturally, the magnons. At the same time the present understanding of the magnon kinetics is far from being perfect. The most information concerning magnons is related to static properties like their contribution to specific heat, etc. In particular, there are only few publications concerning coupling of low frequency magnons to electrons since due to conservation  laws such coupling is suppressed, see, e.\,g., Ref.~\cite{Mills71}.
However, the momentum conservation law in the direction normal to interface is violated for relatively thin ferromagnetic layers  thus allowing efficient electron-magnon coupling down to low magnon frequencies~\cite{KozubCaro}.  To the best of our knowledge,  an experimental  information concerning electron-magnon interactions is far  from being complete. To some extent, this is because of lacking of experimental methods allowing to single out electron-magnon interaction in sufficient detail.
In our opinion,
 useful information can be obtained by studies of magnon-mediated electron drag in a ferromagnetic bilayer since it is the elementary electron-magnon processes that are responsible for such a drag. We believe that the observation of such a drag could give valuable information not only about the efficiency of electron-magnon interactions, but also concerning more delicate details of magnon kinetics like the effect of magnetic domain structure. One notes, in particular, that the domain structure suppressing to some extent the magnon transport also suppresses the drag effect mentioned above. Thus the rearrangement of the domain structure by external magnetic field can give an instrument to separate the drag contribution.

 There exists extensive literature on electron drag in bilayers, see~\cite{RevModPhys.88.025003} and references therein  for a review. Most attention is paid to various structures based on semiconductors, graphene, etc.  However, we are not aware of works aimed on electron drag between typical metals. Indeed, the direct Coulomb drag between typical metals is expected to be very small because of significant screening of electromagnetic fluctuations. Therefore, one can expect that only indirect drag caused by momentum transfer between different quasiparticles can be observed.

\begin{figure}[h]
\centering
\includegraphics[width=6cm]{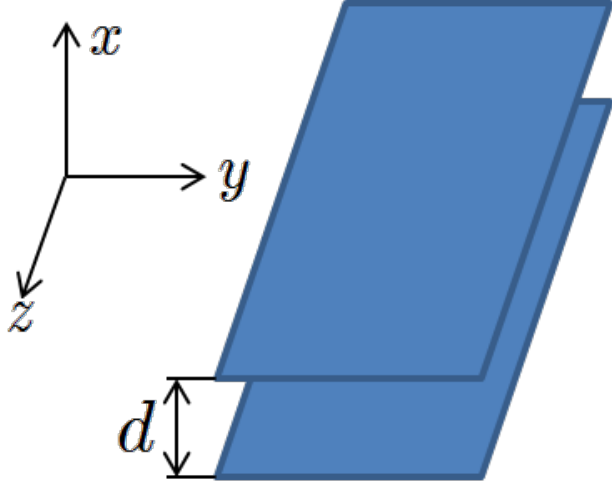}
\caption{(Color online) Geometry of the system. \label{fig1}}
\end{figure}

In this paper, we will consider transconductance of two ferromagnetic layers separated by an insulating interface (possibly - by vacuum gap), see sketch in Fig.~\ref{fig1}. The source of transconductance is expected to be  magnon-electron interactions. Namely, we assume that the external voltage is applied to the ``active" layer stimulating electric current through this layer. In its turn, the scattering of the current-carrying electrons by magnons leads to a magnon drag current within this layer. However the direct effect of this magnon current on the electron current in the second -- ``analyzing" -- layer is negligible. The main contribution to the electrical current is given by the non-equilibrium magnons generated in the analyzing layer due to magnon-magnon interaction. In this paper we assume that the coupling between the two layers is supported by 3-magnon processes. As  known, these processes are supported by dipole-dipole interactions and thus are of non-local nature. Namely, the coupling of magnons from different layers is possible provided the width of the gap is smaller than the magnon wavelength. As a result, the magnon drag current also exists within the ``analyzing layer". In its turn, scattering of these magnons by electrons within the second layer leads to a creation of corresponding drag current.
We will estimate the magnon-induced electron drag with that induced by electron-phonon interaction and find the conditions at which the magnetic effects dominate.

The paper is organized as follows. In Sec.~\ref{coupling} we will evaluate electron-magnon coupling in a thin layer. Based on this evaluation we deribe the transport equations for interacting electrons and magnons in Sec.~\ref{transport}. Electron drag due to coupling with magnons and phonons is considered in Sec.~\ref{drag}. The results are discussed in Sec.~\ref{discussion}. Some detailes of the calculation are clarified in~\ref{append}.

\section{Electron-magnon coupling within a thin ferromagnetic layer}\label{coupling}

Following Mills \emph{et al.},~\cite{Mills71} we write the matrix
element for the transition of an electron in initial state $\bf k$
to final state $\bf k'$ due to creation of a magnon in state $\bf
q$ as
\begin{equation}\label{initial}
{{g}}|_{\bf k \rightarrow k' } = J\frac{(\bf s \cdot M)}{sM} \frac{
a^{3/2}}{ {\Omega}^{3/2}}\int d{\bf r} \exp [i({\bf k - k' -
q})\cdot {\bf r}],
\end{equation}
where $\bf s$ ($s=\sigma\hbar$) and $\bf M$ are the electron spin
and the magnetization, respectively, $J$ is the exchange constant,
$a$ is the lattice constant, while $\Omega$ is the normalizing volume.
 The integration over the infinite sample volume
in Eq.~(\ref{initial}) would  give the standard momentum conservation
law $\bf k=\bf k'+\bf q$. However, for the $x$-direction of the
thin analyzer layer (see Fig.~\ref{fig1}) the integration is performed over the
finite layer thickness $t_a$. This leads to smearing of the momentum conservation law $k_x$. The resulting matrix element renormalized with respect to the momentum non-conservation due the finite thickness is
denoted as $\tilde g$.

We assume that standard momentum conservation holds in plane of
the layer, so that in Eq.~(\ref{initial}) we can concentrate on
the integration over $x$. The corresponding factor arising in the
expression for ${|{\tilde g}|}^2$ readily can be written as
\begin{equation}\label{kx}
(a/t_a) [(k_x -k_x' - q_x) t_a]^{-2} =
a/t_a^3(k_x -k_x' - q_x)^2\, .
\end{equation}
For given initial and final electron energies,  $\varepsilon
=\varepsilon_{\bf k, - \sigma }$ and $\varepsilon'_{\bf k', \sigma
}$, respectively,  the Fermi surfaces are typically separated by a relatively
large gap,  $|\textbf{k}-\textbf{k}'|=\Delta k_F \simeq k_F
E_{\text{ex}}/\varepsilon_F$. For magnons of long wavelength we expect $q
<< \Delta k_F$, which allows neglecting of $q$ in the estimates.
We first integrate over $\vartheta, \vartheta'$ (denoting the
angles of the wave vectors with respect to the $x$-axis) and
$\varphi, \varphi'$ (denoting the angles of the wave vectors with
respect to their in-plane component). Since the difference
$\varepsilon - \varepsilon'$ (controlled by the distribution
functions) is much less than the exchange energy, $E_{\text{ex}}$, we also will neglect this
difference in course of the angular integrations. Thus one has
$k_x = k_{F,-}\cos\vartheta$ and $k_x' = k_{F,+} \cos \vartheta'$.
Momentum conservation in the
$yz$ plane of the layer leads to the relationship
$$\cos \vartheta' = \left[1 - (k_{F,-}/k_{F,+})^2 \sin^2 \vartheta
\right]^{1/2}. $$
Since $k_{F,+}$ is larger than $k_{F,-}$, there clearly is a gap
preventing small values of $\cos \vartheta'$. Further, one obtains
for the denominator of Eq. (\ref{kx}),
$t_a^3k_F^2\left(\xi - \sqrt{\xi^2+2\delta}\right)$,
where $\xi \equiv  \cos \vartheta$.
After integration over $\theta$ we arrive at the estimate
$ a/t_a^3 ( \Delta k_F)^2$ where $\Delta k_F \equiv k_{F,+}- k_{F,+}$.
Finally, after these manipulations ${|{\tilde g}|}^2$ can be
estimated as
\begin{equation}
{|{\tilde g}|}^2 = J^2 \frac  {k_F}{(\Delta k_F)^2t_a}.
\end{equation}

\section{Transport equations for magnons coupled to mobile electrons} \label{transport}

We start from the generic kinetic equation for bosons interacting with electrons, cf. with book \cite{pitaevskii2012physical}, \S 79, Eq.~(79.3):
\begin{equation}
\dot{N}_\bq =I_{\text{coll}}\{N_\bq\}, \label{ke}
\end{equation}
where
\begin{eqnarray}
 &&I_{\text{coll}}= \int\frac{2d^3p}{(2\pi)^3} w(\bp;\bp',\bq) \left[ f_{\bp}(1- f_{\bp'}) (1+N_\bq)
-  f_{\bp'}(1- f_{\bp}) N_\bq \right]
\nonumber \\ && \qquad \qquad\quad  \times
\delta(\varepsilon_{\bp'} + \omega_\bq- \varepsilon_{\bp})\, , \\ &&
w(\bp;\bp',\bq)   = w_0(\bq)\, \delta(\bp'-\bp +\bq). \label{coll1}
\end{eqnarray}
Here $N_\bq$ is the  number of bosons.
This is 3D equation, it is also written for a spin-degenerate case. Let us restore spins since for our case exchange energy matters. We have to replace $2 \to \sum_\sigma$ and then introduce spin indices. We obtain
\begin{eqnarray}\label{te01}
&& I_{coll}= \sum_{\sigma= \pm 1} \int\frac{d^3p}{(2\pi)^3} w(\bp;\bp -\bq,\bq)
\nonumber \\ && \times \left[ f_{\bp,\sigma}(1- f_{\bp',-\sigma}) (1+N_\bq)-  f_{\bp',\sigma}(1- f_{\bp, -\sigma}) N_\bq \right]
\delta(\varepsilon_{\bp'} + \omega_\bq - \varepsilon_{\bp}). \label{coll2}
\end{eqnarray}
To take into account that both magnetic layers are thin we replace
 $w \propto \delta(\bp'-\bp +\bq)$ by
 \begin{eqnarray}\label{tmp1}
&& |\tilde g |^2 \propto \frac{a \mathcal{A} }{{t_a}^3(k_x -k_x' - q_x)^2}
\nonumber \\&& \quad \times \delta (k \sin \theta \sin \varphi - k' \sin \theta' \sin \varphi')\cdot \delta ( k \sin \theta \cos \varphi - k' \sin \theta' \cos \varphi').
\end{eqnarray}
It allows to express the collision operator as the integral over energies,
\begin{eqnarray}\label{rate1ab}
&& \frac{d N_\omega }{d t} = \frac{1}{2 \pi \hbar}
\sum_{\sigma} \int d \varepsilon\, D (\varepsilon) \int d
\varepsilon' \, D(\varepsilon')|{{{\tilde g}}}|^2
\nonumber\\&& \quad \times
 \left[ f_{\epsilon'_{-\sigma} +\omega} (1- f_{\varepsilon'_{-\sigma}})(1+N_\omega) - f_{\epsilon_{\sigma} +\omega} (1- f_{\varepsilon_{\sigma}}) N_\omega\right]\delta (\varepsilon'_{-\sigma}- \varepsilon_{\sigma}).
\end{eqnarray}
Here we have taken into account that the magnon frequency is much smaller than the exchange energy.
See some details of the calculation are given in Appendix A.

We have to take into account that non-equilibrium distribution,  $f$, depends both on the absolute value of $p=\hbar k$, i.e., on energy, and on the direction of $\bp$ in the $y,z$-plane. We can put
\begin{equation}
f(\varepsilon,\sigma,\varphi)=f^{(0)}_{\varepsilon,\sigma}+f^{(1)}_{\varepsilon,\sigma} \sin \theta \cos \varphi = f^{(0)}_{\varepsilon_\sigma}+f^{(1)}_{\varepsilon_\sigma}n_E, \  n_E \equiv (p_E/p) \, .
\end{equation}
Here $p_E $ is the projection of $\bp$ on the electric field.
We are interested in the part proportional to $f^{(1)}_{\varepsilon_\sigma, \be_\bp}$, which provides a source for the magnon drag. For 3D case we would obtain  the contributions
\begin{eqnarray}
 &&\left[f^{(1)}_{\bk,\sigma}(1 - f^{(0)}_{{\bf  k + q}, -\sigma})- f^{(0)}_{{\bf k},  \sigma} f^{(1)}_{{\bf  k + q}, -\sigma})\right]
N^{(0)}_{ {\bf q}}
\nonumber \\ && \qquad \qquad
-\left[(1 - f^{(0)}_{{\bf k}, \sigma})f^{(1)}_{{\bf k + q}, - \sigma} - f^{(1)}_{{\bf k}, \sigma}f^{(0)}_{{\bf k + q}, - \sigma}
\right]     (N_{{\bf q}}^{(0)} +1 )
\nonumber \\ &&  \qquad \qquad \qquad
=  f^{(1)}_{{\bf k}, \sigma} \mathcal{G}_{\mathbf{k+q},-\sigma} - f^{(1)}_{{\bf  k+q}, -\sigma}\mathcal{G}_{\mathbf{k},\sigma},
\end{eqnarray}
where
\begin{equation}
 \mathcal{G}_{\mathbf{k},\sigma} \equiv \left[
f^{(0)}_{{\bf k}, \sigma} (N_{\mathbf{q}}^{(0)}+1)+(1 - f^{(0)}_{{\bf k}, \sigma}) N_{\mathbf{q}}^{(0)}
\right].
\end{equation}

Now we take into account that
\begin{eqnarray}
 f^{(1)}_{{\bf k}, \sigma}&=&\frac{e\tau \hbar \mathbf{k \cdot E}}{m} \left( -\frac{\partial f^{(0)}(\varepsilon)}{\partial \varepsilon}\right)_{\varepsilon=\varepsilon_{\mathbf{k},\sigma}} ,
 \nonumber \\
  f^{(1)}_{{\bf k+q}, -\sigma}&=&\frac{e\tau \hbar \mathbf{(k+q) \cdot E}}{m} \left( -\frac{\partial f^{(0)}(\varepsilon)}{\partial \varepsilon}\right)_{ \varepsilon=\varepsilon_{\mathbf{k},\sigma}+\hbar \omega} . \nonumber
\end{eqnarray}
One notices that
\begin{eqnarray*}
\mathcal{G}_{\mathbf{k+q},-\sigma} \left( -\frac{\partial f^{(0)}(\varepsilon)}{\partial \varepsilon}\right)_{\varepsilon = \varepsilon_{\mathbf{k},\sigma}}
&=& \mathcal{G}_{\mathbf{k},\sigma}  \left( -\frac{\partial f^{(0)}(\varepsilon)}{\partial \varepsilon}\right)_{\varepsilon=\varepsilon_{\mathbf{k},\sigma}+\hbar \omega}
\\
&\approx & \frac{\delta(\varepsilon_{\mathbf{k},\sigma}-\mu)}{\sinh (\beta \omega_\mathbf{q})}.
\end{eqnarray*}
Now we can choose one of the axes in the $yz$-plane along the electric field. We can write
$$  f^{(1)}_{{\bf k}, \sigma} \mathcal{G}_{\mathbf{k+q},-\sigma} - f^{(1)}_{{\bf  k+q}, -\sigma}\mathcal{G}_{\mathbf{k},\sigma}\approx\frac{e\tau \hbar (\mathbf{q\cdot E})}{m\sinh (\beta \omega_\bq)}\delta(\varepsilon_\sigma - \mu). $$
Now we can come back to Eq.~(\ref{rate1ab}) and write the electronic part as
\begin{eqnarray}
- \frac{(\bq\cdot \mathbf{E})}{  \sinh (\beta \omega_{\mathbf{q}})}
\frac{e \hbar \tau |\tilde{g}|^2 }{ m }\sum_\sigma D_\sigma (\mu)D_{-\sigma}(\mu).
\end{eqnarray}
Here $ D_\sigma (\mu)$ is the partial electron density at the Fermi level $\mu$.
Since the collision operator linear in $\delta N_\bq = N_\bq- N_\bq^{(0)}$ can be expressed as $-\delta N_\bq/\tau_m$ where
\begin{equation} \label{tau_m}
\frac{1}{\tau_m}  =\frac{1}{\tau_{me}}+ \frac{1}{\tau_{mm}} + \frac{1}{\tau_{mb}}
\end{equation}
where the partial contributions are due to magnon-electron magnon-magnon and magnon-background  scattering, respectively.

Therefore, the source for the drag, which is odd in $\bq$ can be expressed as
\begin{equation} \label{deltaN}
\delta N_{\mathbf{q}} =- \frac{(\mathbf{q}\cdot \mathbf{E}) |\tilde{g}|^2}{  \sinh (\beta \omega_{\mathbf{q}})}
\frac{e \hbar \tau \tau_m }{ m }\sum_\sigma D_\sigma(\mu) D_{-\sigma} (\mu).
\end{equation}
Taking into account that  $\tau_{me}^{-1}\simeq \hbar^{-1}\left|
\widetilde{g}\right| ^{2}[D(\epsilon_F)]^2\hbar
\omega_{\mathbf{q}}$ (see \cite{KozubCaro}) one obtains
$$\delta N_\bq \propto  (\bq\cdot \mathbf{E}) \frac{\tau_e \tau_m}{\tau_{me}}. $$
Estimating the ratio $\tau_m/\tau_{me} \approx 1$ { and taking into account that for thermal phonons $\sinh (\beta \omega_{\mathbf{q}})\sim 1$
we get
\begin{equation}\label{magdrag}
\delta N_\bq \approx \frac{e\tau_e \hbar (\bq \cdot \bE)}{m \varepsilon_F} .
\end{equation}

\section{Drag in a system of two magnetic layers} \label{drag}

Now let us consider a coupling of two ferromagnetic layers where the one supports a current flow while the second (separated from the first one, say, by vacuum gap) is the "analyzer" where a drag current can be produced by magnons created in this layer. First, we assume that the direct coupling between the magnons in the first layer and electrons within the ``analyzer" layer can be neglected since the coupling constant given by Eq.~(\ref{initial}) implies exchange interaction dramatically decreasing with distance.
Therefore, we take into account only 3-magnon processes based on the dipole-dipole interaction.  The latter allows a gap between the layers with a thickness less than the magnon wavelength.

In our case we can consider the magnon-magnon interactions like in the case of 3D geometry since the magnon wavelength for magnons with energies of the order of several K is much smaller than the thickness of the layer ($\gtrsim 10$~nm).
 Thus, according to Ref.~\cite{Akhiezer}  for processes involving thermal magnons we have an estimate
\begin{equation} \label{3ma}
 \frac{1}{\tau_{mm}} \sim \frac{2\pi}{\hbar}\frac{(\mu_B M)^2 T^{1/2}}{
\Theta_C^{3/2}}.
\end{equation}
Here $\Theta_C$ is the Curie temperature, $M$ is the magnetization, $\mu_B$ is Bohr magneton. We have taken into account that the magnon spectrum can be represented as
\begin{equation}
\hbar \omega_\bk = \varepsilon_0 + \Theta_C (ak)^2,
\end{equation}
where  $\varepsilon_0 \sim 2 \mu_B (H + M)$, $H$ is external magnetic field.
We have taken into account that for thermal magnons $(ka) \sim (T/\Theta_C)^{1/2}$ where $k$ is the typical thermal magnon wave vector while $a$ is the lattice constant.
Putting $\mu_0 M \sim 1$~T where $\mu_0$ is the vacuum permeability,  $\Theta_C = 10^3$~K, $q_2 \sim q_3 = (m\omega_3/\hbar)^{1/2}$, $\hbar \omega_2\sim \hbar \omega_3 \sim \hbar \omega_1 \sim T \sim 10 K$,  one obtains
\begin{equation}\label{t3m}
\tau_{mm}^{-1}\approx  \times 10^8 \  \mathrm{s}^{-1}.
\end{equation}
Then, assuming that while the effect of magnons from the first layer on the magnon system of the second layer is
$\propto 1/\tau_{mm}$ while the relaxation of the non-equilibrium magnon distribution in the second layer is supported, mostly, by electron-magnon processes with a rate~\cite{KozubCaro},
\begin{equation} \label{tme}
\tau_{me}^{-1}\simeq \hbar^{-1}\left|
\widetilde{g}\right| ^{2}[D(\epsilon_F)]^2\hbar
\omega_{\mathbf{q}}
\end{equation}
and, second, by other background scattering mechanisms characterized by the relaxation time $\tau_{mb}$,
one can conclude that the contribution to the magnon drag in the passive layer can be estimated  as
\begin{equation}\label{magdrag1}
{N}_d = \delta N_\bq \frac{ \tau_{me}}{\tau_{mm}}= \frac{e\tau_e \hbar ({\bf q E})}{m\varepsilon_F} N_0 (\hbar \omega_\bq) \frac{\tau_{me}}{\tau_{mm}} .
\end{equation}
Here ${N}_d(\bq)$ is  the drag component of the magnon distribution in the passive layer, while $\delta N_\bq$ is the non-equilibrium addition to magnon distribution in the active layer, Eq.~(\ref{magdrag}),  while for thermal magnons $N_0(T)  \sim 1$. This estimate follows from a simple rate equation $N_d/\tau_{me} = \delta N_\bq/\tau_{mm}$
where r.h.s. is the source term (magnon-magnon collision operator describing relaxation of non-equilibrium function $\delta N_\bq$ which gives the ``in" term in collision integral for $N_d$ while we assume that the magnons are mainly scattered by electrons. We will estimate the ratio  $\tau_{me}/\tau_{mm}$ later.

Now we are able to estimate the drag contribution to the electron current within the second layer. We start be expressing the drag current through the electron-magnon collision operator, $I_{em}$:
\begin{equation}\label{curr1}
J_d = e\sum_{\bf p} v \tau I_{em}.
\end{equation}
Here $\tau$ is electron relaxation time. One notes that the contribution to $N_{\bf q}$ which is even in ${\bf q}$ does not lead to any contribution to $f_{\bf k}$ which would be odd in $\bf k$. It is only the contribution odd in $\bf q$, $N_d$, which leads to current-carrying contribution to $f$.

As it was noted earlier, we are interested to compare the magnon contribution to the drag in the ``passive'' layer with the effect resulting from non-equilibrium magnons, generated in the ``active" layer and penetrating to the "passive" layer. Thus before the detailed estimate of the magnon-induced effect,
let us first consider the well known case of phonon-induced drag. After that we will specify the differences for the case of magnons.
After rather standard analysis one can express the electron-phonon collision integral as
\begin{eqnarray}\label{eq3}
&& I_{e-ph} = \int \frac{d^3 q}{(2\pi)^3}\int dx d \varphi \,
 | C|^2
(f_{\bf k} - f_{\bf k + q}) N_{d,ph} (\bq)
\nonumber\\ && \quad \times
 \left[\delta(\varepsilon_{\bf k + q} - \varepsilon +
\hbar \omega) - \delta(\varepsilon_{\bf k + q} - \varepsilon -
\hbar \omega)\right]
\end{eqnarray}
where $x = \cos \theta$, $\theta \equiv \angle \bq\bp$, $N_{d,ph} \propto ({\bf q}{\bf E})$ is the field-induced addition to phonon
distribution function.
One has in mind that the brackets with delta-functions can be rewritten as
\begin{equation*}
\frac{m}{p\hbar q} \left[\delta \left( \cos\theta  + \frac{\hbar \omega m}{p \hbar q}\right) - \delta \left( \cos \theta  - \frac{\hbar \omega m}{p \hbar q}\right)\right].
\end{equation*}
We have also taken into account that $ N_d \propto (\bq \cdot \bE)$.
Denoting $\angle {\bf q E} = \alpha $ and $\angle {\bf E p} = \phi$ and using the relationship
\begin{equation*}
\cos \alpha = \cos \theta \cos \phi + \sin \theta \sin \phi \cos \varphi
\end{equation*}
where $\varphi$ is an angle between the planes ${\bf qE}$ and ${\bf Ep}$
one can cast $I_{e-ph}$ into the form:
\begin{eqnarray*}
&& \int\frac{q^2 dq \, x^2dx}{(2\pi)^3}|C|^2 \tilde N_{d,ph} \hbar qv \cos \phi \frac{m}{\hbar pq} \frac{\partial f_0}{\partial \varepsilon} \\ && \quad \times
\left[\delta \left( x + \frac{\hbar \omega m}{p \hbar q}\right) - \delta \left( x - \frac{\hbar \omega m}{p \hbar q}\right)\right]
\end{eqnarray*}
where $\tilde N_{d,ph}$ denotes the factor resulting from $N_{d,ph}$ where the $\cos \theta$-factor is replaced by 1.
Finally one arrives at the estimate
\begin{equation}\label{driftph}
I_{e-ph} \sim
\frac{ 2|C(q_T)|^2 }{(2 \pi)^3} {\tilde N_d(q_T)}\cos \phi \frac{\partial f_0}{\partial \varepsilon}  \left(\frac{\omega m}{p }\right)^3.
\end{equation}
Now we take into account that $|C(q_T)|^2q^3m/(p\hbar q) \simeq \tau_{e-ph}^{-1}$ and the fact that integration over $q$ is restricted by temperature.
We also assume that the phonon distribution is completely controlled by electrons and thus the drift velocities of electrons and phonons have the same order of magnitude,  $ e\tau E/m$.
Thus the non-equilibrium contribution to the phonon distribution function is of the order of
$$\tilde N_d \sim N_0 (T/\hbar)  (e\tau E/m w) \sim (e\tau E/m w)$$ where $N_0(\omega_\bq)$ is the equilibrium distribution
while $w$ is the sound velocity. Neglecting the phonon interface scattering and using Eqs.~(\ref{curr1}) and (\ref{driftph}) one obtains:
\begin{equation}
j_d = j_a\left(\frac{w}{v_F}\right)^2\frac{\tau_e}{\tau_{e-ph}}\frac{\hbar q}{p_{F}} \propto T^4
\end{equation}
where $j_a= (e^2\tau_e n/m) E $ is the current in the active layer.

Now let us return to the magnons. Actually we are only interested in contribution to the magnon distribution function which is odd in wave vector in-plane component. Producing derivations which are similar to ones which are applied to calculate phonon drag starting from electron-phonon collision operator we obtain the corresponding contribution to electron-magnon operator:
\begin{eqnarray}\label{eq10}
&&I_{em} = \int \! \!  q_{\parallel}  \! \!   d q_{\parallel}\int  \! \!   d q_{\perp} \! \!   \int   \! \!  d  k_{\perp}
 |{\tilde g}|^2  N_d(\omega, \bq_{\parallel}) A(\bk,\bq, \sigma) ,\ \label{iem1} \\
 &&A\equiv (f_{\sigma, k_{\parallel}, k_{\perp}} - f_{-\sigma, k_{\parallel} + q, k'_{\perp}})
\nonumber  \\ && \quad  \times
\left[\delta ( \varepsilon_{\sigma, k_{\parallel},  k_{\perp}}
-  \varepsilon_{-\sigma, k_{\parallel} + q,  k'_{\perp}} + \hbar \omega )
 \right. \nonumber \\ &&\quad \quad \quad  \left.- \delta ( \varepsilon_{\sigma, k_{\parallel},  k_{\perp}}  -  \varepsilon_{-\sigma, k_{\parallel} + q,  k`_{\perp}} -  \hbar \omega )\right]. \label{A}
\end{eqnarray}
Here we neglect the component $q_x$  due  to non-conservation of the momentum in normal (to the interface) direction, as we have done in course of estimates  of the electron-magnon matrix element.
Thus we have
\begin{equation*}
\hbar \omega \frac{\partial f}{\partial \varepsilon}  \delta ( \varepsilon_{\sigma, k_{\parallel},  k_{\perp}}  -  \varepsilon_{-\sigma, k_{\parallel} + q,  k'_{\perp}}  ).
\end{equation*}
As a result of integration over $k'_{\perp}$ we get
\begin{equation*}
E_{ex}  - { \hbar \bf q_{\parallel}  v_{\parallel}}  = ( \hbar k_{\perp})^2/(2m).
\end{equation*}
While the initial derivation of the matrix element $g$ did not take into account any asymmetry of the magnon spectrum, now we see that actually such a dependence exists since the corresponding matrix element depends on the value  $k_{\perp}$.
We can estimate the variation of the matrix element due to asymmetry of the magnon spectrum as
\begin{equation}
|g^2| \sim |g^2(0)|( 1 +  \hbar \bq_{\parallel}  \bv_{\parallel}/E_{ex}).
\end{equation}
While without this $q$-dependent correction the integration over $q_{\parallel}$ of the integrand including $N_d$ would vanish, this correction supports the drag effect.

Let us recall Eq.~(\ref{magdrag1}) for the non-equilibrium magnons distribution in the passive layer.
Specifying contribution of drag contribution to $I_{em}$ we have
\begin{equation}
\delta I_{em} \sim \frac{e \tau_e}{\tau_{em}}\frac{({\bf v E})(\hbar q)^2}{m \varepsilon_F E_{ex}}\frac{\tau_{me}}{\tau_{mm}}\hbar \omega _\bq\frac{\partial f_0}{\partial \varepsilon}.
\end{equation}
Now we shall specify the ratio $\tau_{me}/\tau_{em}$. With a help of definition of $\tau_{me}$ and taking into account that the estimate for $\tau_{em}$ follows from substitution to $I_{em}$ of equilibrium magnon function $N_0$ (accounting also for the fact that the delta-function in this case practically are not sensitive to magnon wave vector since are controlled mostly by the component normal to the interface)
\begin{eqnarray}
&& \tau_{em} \sim (q_T a)^3 D_{\varepsilon} \frac{|g^2|}{\hbar} \sim  \frac{(\hbar q_T)^2}{m_m}\frac{m_m (q_T a) a^2}{\hbar^2} D_{\varepsilon} \frac{|g^2|}{\hbar}
\nonumber\\  &&   \qquad \to
\frac{\tau_{me}}{\tau_{em}} \sim   \frac{\varepsilon_F}{E_{ex}} \left(\frac{T}{E_{ex}}\right)^{1/2}.
\end{eqnarray}
Here we have taken into account that the integration over $q$ is restricted by temperature while $m_m  \sim  a (2E_{ex})^{1/2}/\hbar$ is magnon mass (by an order of magnitude larger that the electron mass and $(\hbar q)^2/2 m_m = \hbar \omega_\bq $ is the magnon energy.
Integration of $I_{em}$ over $\bf p$ with a weight $e\bf v$ (which gives the drag current) yields
\begin{equation}\label{magdrag}
j_d =  j_a  \frac{m_m}{m}\left(\frac{T}{E_{ex}}\right)^{5/2} \frac{\tau_e}{\tau_{mm}}
\end{equation}
Here  $m$ is the electron effective mass. Putting $m_m/m \sim 10$, $T =10$~K, $E_{ex}/k_B \sim 10^3$~K, $\tau_e \sim 10^{-13}$~s, $\tau_{mm}\sim 10^{-7}$~s we obtain $(j_d/j_a) \sim 10^{-8}$. Though the effect is small it seems to be observable.

Thus the relation between magnon and phonon contribution is given as
\begin{equation*}
\frac{m_m}{m}\left(\frac{T}{E_{ex}}\right)^{5/2}\frac{\tau_e}{\tau_{mm}}
\left[\frac{w}{v_F}\frac{\tau_e}{\tau_{e-ph}}\frac{T}{\varepsilon_F} \right]^{-1}.
\end{equation*}
Consequently,
 for ideal mechanical contact between the layers (as it was suggested before), the magnon contribution dominates provided
\begin{equation*}
\frac{m_m}{m}\left(\frac{T}{E_{ex}}\right)^{5/2}\frac{\tau_e}{\tau_{mm}} > \frac{w}{v_F}\frac{\tau_e}{\tau_{e-ph}}\frac{T}{\varepsilon_F},
 \end{equation*}
 which can be rewritten as
 \begin{equation}\label{ph-m}
 \left(\frac{T}{E_{ex}}\right)^{3/2}\frac{m_m}{m} > \frac{\tau_{mm}}{\tau_{e-ph}}\frac{w}{v_F}\frac{E_{ex}}{\varepsilon_F}.
 \end{equation}

 Having in mind an estimate $\tau_{mm}^{-1}\sim 2\cdot 10^7$~s$^{-1}$ at $T = 1$ K we note that the value of $\tau_{e-ph}$ for the same temperatures gives nearly the same estimate. The ratio $m_m/m \sim 10$, $E_{ex}/\varepsilon_F \sim 0.1$,  while $w/v \sim 10^{-3}$. Thus the r.h.s. of Eq.~(\ref{ph-m}) is of the order of $10^{-4}$. At the same time, l.h.s. at $T =
 1~K$ is of the order of $10^{-3}$. Thus one concludes that at low temperatures the magnon contribution can dominate the phonon one. However at $T = 10~K$ for ideal acoustic contacts within the structure the phonon contribution appears to be somewhat (by a factor of 3) bigger than the magnon contribution. Nevertheless one expects an additional factor in favor of the magnon contribution due to acoustic mismatch between the ferromagnetic layers and the interlayer. The mismatch would suppress the phonon contribution. We also note that effects of magnetic field discussed in the next section give additional mechanism to separate the magnon contribution.

\section{Discussion and conclusions} \label{discussion}

Now let us return to the estimate of the drag current given by Eq.~(\ref{magdrag}). As it is seen, the result is independent of the electron-magnon coupling constant ${\tilde g}$. This result seems to be in contradiction to our suggestion to use the drag effect for studies of electron-magnon interactions. However, here we would like to note that the fact that the sensitivity of the drag effect to the coupling constant  $\tilde g$ was washed out mainly due to our neglecting any mechanisms of magnon relaxation except the magnon-electron one. Indeed, in this case the non-equilibrium magnon distribution in  active  layer is, on the one hand, driven by electron-magnon interactions involving non-equilibrium electron distribution. On the other hand, the non-equilibrium magnon distribution relaxes due to magnons scattering by equilibrium electrons. Thus the constant $\tilde g$ is canceled. In its turn, in the  passive  layer the momentum transfer to magnons takes place due to non-equilibrium magnons in the  active  layer, however the corresponding non-equilibrium addition the the distribution of magnons in the passive layer relaxes, again, due to equilibrium electron distribution in the  passive  layer. Thus this addition appears to be proportional to magnon-electron relaxation time. At the same time the drag of electrons by magnons is naturally proportional to the rate of {\it electron-magnon} relaxation. Thus the final effect appears to be proportional to the ratio $\tau_{me}/\tau_{em}$ which does not depend on $\tilde g$.
While being consistent for the case when any relaxations are dominated by electron-magnon coupling, this picture does not hold when some other factors affect the relaxation of the magnon momentum rather than electron contribution. In particular, an important mechanism of such a relaxation can be related to the domain walls. Indeed, the magnons are expected to be scattered by any inhomogeneity  of the magnetization within the sample, domain wall being the typical example of such inhomogeneity. Since the momentum relaxation of magnons in this case can be estimated as $\tau_{m}^{-1} = \tau_{me}^{-1} + \tau_{dw}$ (where $\tau_{dw}$ is related to contribution of the domain walls), then $\tau_m$ entering Eq.~(\ref{deltaN}) is completely controlled by $\tau_{dw}$ provided $\tau_{dw } < \tau_{me}$.  The value of $\tau_{dw}$ can be very roughly estimated as $\tau_{dw} \sim L_d/v_m$ where $L_d$ is maximal distance between the two domain walls (related to the domain with direction along the external magnetic field if the latter is applied) while $v_m \sim ( 2 \hbar \omega/m_m )^{1/2}$ is the magnon velocity. It is important that the value $\tau_{dw}$ appears to be sensitive to the applied magnetic field $H$ affecting the value of $L_d$. The corresponding behavior  can be a delicate one depending on the character of the domain structure. For the thin films which are considered in this paper the most natural domain pattern is related to stripe domains, see their visualization using magneto-optical imaging~\cite{Johansen}.
 According to the estimates obtained in that paper for a typical ferromagnetic material, $L_d $, is of the order of the film thickness at  $H = 0$ and diverges when $H \rightarrow H_c$ (as $(H_c  -H)^{-1/2}$ where $H_c$ corresponds to infinite period of the domain structure (when external field is close to the saturation field).  We appreciate that the picture considered in the paper~\cite{Johansen} is rather a model one and at least does not take into account a presence of the second ferromagnetic layer. Nevertheless we believe that our considerations can give at least semi-quantitative estimates. Namely, if at $H = 0$ the drag effect is controlled by the domain walls (since at $H = 0$ the width of the domains is minimal),  it starts to increase with an increase of $H$ until the moment when magnon-electron interaction starts to dominate which takes place when $L_d \simeq v_m \tau_{me}$.  Correspondingly, the drag effect is saturated. It is this saturation point which allows in principle to estimate an efficiency of magnon-electron interactions.

Another effect of external magnetic field can be related to mutual orientation of the magnetic field and the driving current in the ``active" plane. One expects that, since the domain structure is expected to be oriented along the direction of external field, the drag effect is more pronounced when the direction of the driving current is along the direction of magnetic field. Indeed, in this case both magnon and electron transport suffer much less effect of the domain walls than if the current would be directed normally to the domain walls. Correspondingly, a pronounced anisotropy of the effect with respect to the driving current direction is expected.

As for temperature dependence of the effect, one can only conclude that the effect strongly increases with temperature increase -- according to Eq.~(\ref{magdrag}) proportionally to $T^{5/2}$. However, the phonon contribution has stronger temperature gain ($\propto T^4$). Nevertheless, the phonon effect is not expected to be sensitive to the external field (at least at weak fields which do not affect the resistance). Thus the studies of the magnon drag at higher temperatures can make sense in combination with the effects of external magnetic field.

\section*{Acknowledgment}
This work was supported by the Russian Foundation for Basic
Research (project No. 16-02-00064)

\appendix
\section{Relaxing the conservation law} \label{append}
The first $\delta$-function in Eq.~(\ref{tmp1}) can be rewritten as
$$\frac{1}{|\cos \varphi_1|}\delta (\varphi' -\varphi_1) \ \textrm{where} \
\varphi_1 \left(\frac{k}{k'},\theta,\theta',\varphi \right) = \arcsin\left(\frac{k}{k'}\frac{\sin \theta}{\sin \theta'}\right)\sin \varphi. $$
Similarly, the second $\delta$-function in Eq.~(\ref{tmp1}) can be rewritten as
$$\frac{1}{|\sin \varphi_2|}\delta (\varphi' -\varphi_2) \ \textrm{where} \
\varphi_2 \left(\frac{k}{k'},\theta,\theta',\varphi \right)= \arccos\left(\frac{k}{k'}\frac{\sin \theta}{\sin \theta'}\right)\sin \varphi. $$
As a result, after integration over $\varphi'$ we get
$(4\pi/|\sin 2 \varphi_1 | )\delta(\varphi_1 - \varphi_2)$.
This expression is compatible with the relationship $\cos \theta' = \left[1 - (k/k')^2 \sin^2 \theta
\right]^{1/2}$
 between
$k$, $k'$, $\theta$, and $\theta'$, which can be rewritten as
\begin{equation}
k'\cos\theta =\sqrt{k'^{2} -k^2\sin^2\theta}.
\end{equation}
That can be rewritten as
\begin{eqnarray}
&&k'\cos\theta'  - k\cos\theta=\sqrt{(k'^{2} -k^2) + k^2\cos^2\theta} -k\cos \theta \approx \frac{k'^{2} -k^2}{2k \cos \theta}
\nonumber \\  &&
\approx \frac{2m}{\hbar^2}
\frac{\varepsilon'_{-\sigma} - \varepsilon_{\sigma} +\Delta_\sigma -\Delta_{-\sigma}}{2 k_F \cos \theta}
\approx \frac{2m}{\hbar^2 k_F \cos \theta} (\pm \hbar \omega_q + \Delta _\sigma)
\, .
\end{eqnarray}

Then we perform triple integration
\begin{equation}\label{tre}
\int \int \int d\theta\, d\theta' \, d\varphi \,  \mathcal {F}(k,k',\theta, \theta',\varphi) \,
 \delta\left[\varphi_1(k/k',\theta,\theta',\varphi) - \varphi_2(k/k',\theta,\theta',\varphi)\right]
 \end{equation}
where
 \begin{equation}\label{tre1}
 \mathcal{F} \propto \left[k\cos\theta - k'\cos \theta' -q_x)^2 |\sin 2 \varphi_1 (k/k',\theta,\theta',\varphi)|\right]^{-1}.
 \end{equation}
Since $ \mathcal {F}(k,k') $ is  a smooth function of $k$ and $k'$ we assume that
$\varepsilon_{k,\sigma}=\Delta_\sigma +\hbar^2 k^2/2m$ and continue the analysis.

Now we can replace $(k\cos\theta - k'\cos \theta' -q_x)^2 $ in the denominator by $(k_F\Delta_\sigma / \varepsilon_F \cos \theta)^2$ to obtain the estimate  $ \sim k_F^{-2}( \varepsilon_F/\Delta_\sigma)^2$.
Thus we are left with integration over $k$ and $k'$ which is easily reduced to integration over energies to get Eq.~(\ref{rate1ab}). The triple integral (\ref{tre}) is absorbed by $|\tilde{g}|^2$.

%\bibliography{magnons}

\end{document}